\documentclass[aps]{revtex4}
\usepackage{etoolbox}
\usepackage[colorlinks=true]{hyperref}
\pdfoutput=1
\usepackage{graphics}      
\usepackage{graphicx}      
\usepackage{longtable}     
\usepackage{url}           
\usepackage{bm}            
\usepackage{amsmath}
\usepackage{amssymb}
\usepackage{wasysym}
\usepackage{color,soul}
\usepackage{ulem} 
\usepackage{subfigure}
\usepackage{bbold}

\newcommand{\angstrom}{\textup{\AA}}
\providecommand{\av}[1]{ \langle #1 \rangle } %

\begin{document}
	
\title[]{Melting process of twisted DNA in a thermal bath}
\author{O. Farzadian}
\affiliation{Mechanical and Aerospace Engineering, School of Engineering and Digital Sciences, Nazarbayev University, Nur-Sultan 010000, Kazakhstan}
\email{omid.farzadian@gmail.com}
\author{T. Oikonomou}
\affiliation{College of Engineering and Computer Science, VinUniversity, Vietnam}
\author{M. Moradkhani}
\affiliation{Islamic Azad University, Abhar Branch, Abhar, Iran}

\begin{abstract}
We investigate melting transition of DNA sequences embedded in a Langevin fluctuation-dissipation thermal bath.  Torsional effects are considered by a twist angle $\varphi$ between neighboring base pairs stacked along the molecule backbone. Our simulation results show that the increase of twist angle translates linearly  the melting temperature with a positive slope. After the so called equilibrium angle $\varphi_\mathrm{eq}$, the DNA chain becomes very rigid against opening and accordingly very high temperatures are required to initiate the melting process. In such cases however, the biofunctionality of DNA is destroyed before so that the observed in our model melting process becomes biologically irrelevant. We believe that the outcome of this survey would deeper understanding of the interplay between DNA twisting and melting transition for precise control of DNA behavior.  
\end{abstract}



\eid{ }
\date{\today }
\startpage{1}
\endpage{1}
\maketitle


\section{\label{sec:Int}Introduction}

Deoxyribonucleic acid (DNA) molecule carries the genetic code in terms of the four letter alphabet comprised by four kinds of nucleotides Adenine ($\mathrm{A}$), Thymine ($\mathrm{T}$), Cytosine ($\mathrm{C}$), and Guanine ($\mathrm{G}$) \cite{Nature2004}. Accessing this code is possible through fundamental biological processes such as replication and transcription \cite{Alexandrov2008,Alexandrov2009,Lubelsky2014}. Due to the high degree of complexity of DNA dynamics, the details of transcription and replication are not adequately understood and a satisfactory descriptive model is difficult to design. However, these processes initiate with the formation of locally opening and closing of the double helix, a phenomenon termed $\it{breathing}$ motion. At a certain temperature, or under a special circumstance, this local separation of the two strands extends over the entire molecule resulting in a complete separation of the two strands, a phenomenon known as denaturation or melting. Thus, studying DNA denaturation is, in addition to being very interesting in itself, considered a well-grounded step towards the full comprehension of the mechanisms involved in transcription and
replication.

There are several approaches to figure out the dynamics of DNA and the thermal properties of the double helix including eventually its melting \cite{Peyrard1989,Dauxois1993,Zdravkovic2011,Zoli2011}. The Hamiltonian approach, generally based on the one-dimensional Peyrard-Bishop-Dauxois (PBD) model \cite{Dauxois1993,Peyrard2004}, is one the most successful in this issue. The PBD model cosists of nonlinearity in both inside Morse potential and the stacking potential . Beyond the study of the melting curves, this simple model has been applied in other contexts. For example, it has been used to model the open regions in short DNA hairpins \cite{Ares2005,Peyrard2008} and distributions of bubble Lifetimes and bubble Lengths \cite{Skokos2021,Skokos2020}. One of the most interesting improvements to the model referred to as Barbi-Cocco-Payrard (BCP) model which consists of a new parameter to describe the helicity of DNA \cite{Barbi1999}. Introducing the angle of rotation between a base pair and the previous one in a PB-type model in polar cylindrical coordinates, was a main step to tackle this issue \cite{Barbi1999,Cocco1999,Campa2001,Barbi2003}. 

There are two main objectives of this article. First, we perform Langevin dynamics with the BCP model by examining melting the transition as a function of twist angle and
temperatures. Second, we look at a bubble formation probability and thier lifetime.

The paper is organized as follows: In Sec.~\ref{sec:Model}, we introduce the model and we obtain the equation of motion of DNA nucleotide. In Sec.~\ref{sec:Numerical}, the numerical results of our model are presented. Thermal denaturation of DNA described by bubble life times Sec.~\ref{sec:BF}. Finally, a summary of the results is presented in Sec.~\ref{sec:Dis}.

\section{Lagrangian Model}\label{sec:Model}

We study the dynamics of DNA which takes into account the twist-opening interactions due to the helicoidal molecular geometry. It can describe the melting transition and denaturation bubbles of the double strand DNA such as those that occur during the initial stage of the transcription process \cite{Manghi2016}. 

Our starting point is the model introduced \cite{Barbi1999}. The bases can move only in planes perpendicular to the helix axis; besides, the center of mass of the base pair is held fixed, and the two complementary bases move symmetrically with respect to the axis of the molecule. Then for each base pair there are two degrees of
freedom: $r_n$ is the distance between each one of the complementary bases in the $n$th base pair and the helix axis; $\phi_n$ is
the angle that the line joining the two complementary bases
makes with a given direction in the planes where the bases
move.

In the current model, a DNA chain of $N$ base pairs (bps) is described by the following Lagrangian,
\begin{eqnarray}\label{eq:Lagrangian}
\mathcal{L}&=&m\sum_n(\dot{r}_n^2+r_n^2\dot{\phi}_n^2)^2-\sum_n D_n\left[e^{-\alpha_n (r_n-R_0)}-1\right]^2 -K\sum_n(L_{n,n-1}-L_0)^2 \nonumber \\
&&
-S\sum_n e^{-\beta(r_n+r_{n-1}-2R_0)}(r_n-r_{n-1})^2,
\end{eqnarray}
where overdots represent the time derivative.

The first term in the Lagrangian is the kinetic energy. The second term is Morse on-site potential which intended to describe the hydrogen bond interaction between the two bases in a pairs between the two strands. Nucleotides on each strand are attached to the nucleotides of the other strand uniquely as $\mathrm{CG}$ and $\mathrm{AT}$. The parameters $D_n$ and $\alpha_n$ denote the dissociation energy and the inverse length which sets the potential range, respectively, and depend on the nature of the $n$th base pair $\mathrm{AT}/\mathrm{CG}$. $\mathrm{C}$ and $\mathrm{G}$ ($\mathrm{A}$ and $\mathrm{T}$) are bound together by three (two) hydrogen bonds with $D_{\mathrm{CG}} = 1.5 D_{\mathrm{AT}}$ and $\alpha_{\mathrm{CG}} = 1.64\alpha_{\mathrm{AT}}$. Also, equilibrium distance between base pairs (equilibrium value of $r_n$) is $R_0=10 \angstrom$.

The quadratic term in $(L_{n,n-1}-L_0)^2$, represents the elastic energy of the backbone rods between neighboring base-pairs on each strand. $L_{n,n-1}$ is the axial distance between successive base pair planes on the same strand, and as a function of $\varphi_n=\phi_n-\phi_{n-1}$ is given by 
\begin{eqnarray}\label{eq:actualdistance}
L_{n,n-1}=\sqrt{h^2+r_n^2+r_{n-1}^2 - 2 r_n r_{n-1} \cos(\varphi_n)} \,,
\end{eqnarray}
where $h=3.4\angstrom$ is the fixed distance between neighbor base
planes.
$L_0$ is the same function computed for equilibrium configuration, $r_n=r_{n-1}=R_0$ and $\varphi_n=\varphi_\mathrm{eq}=2\pi/10.4\approx 34.6^\circ$ (approximately 10 base pairs per helix turn \cite{Wang1979}).

Finally, the last term models a stacking interaction
between neighboring base pairs. Its effect is to decrease the
stiffness of the open parts of the chain relatively to the closed
ones and to stabilize the latter with respect to the denaturation
of a single base-pair. Terms of this type increase the
cooperative effects close to the melting transition. 
\begin{figure}[hbtp]
	\centering
	\includegraphics[width=0.6\textwidth]{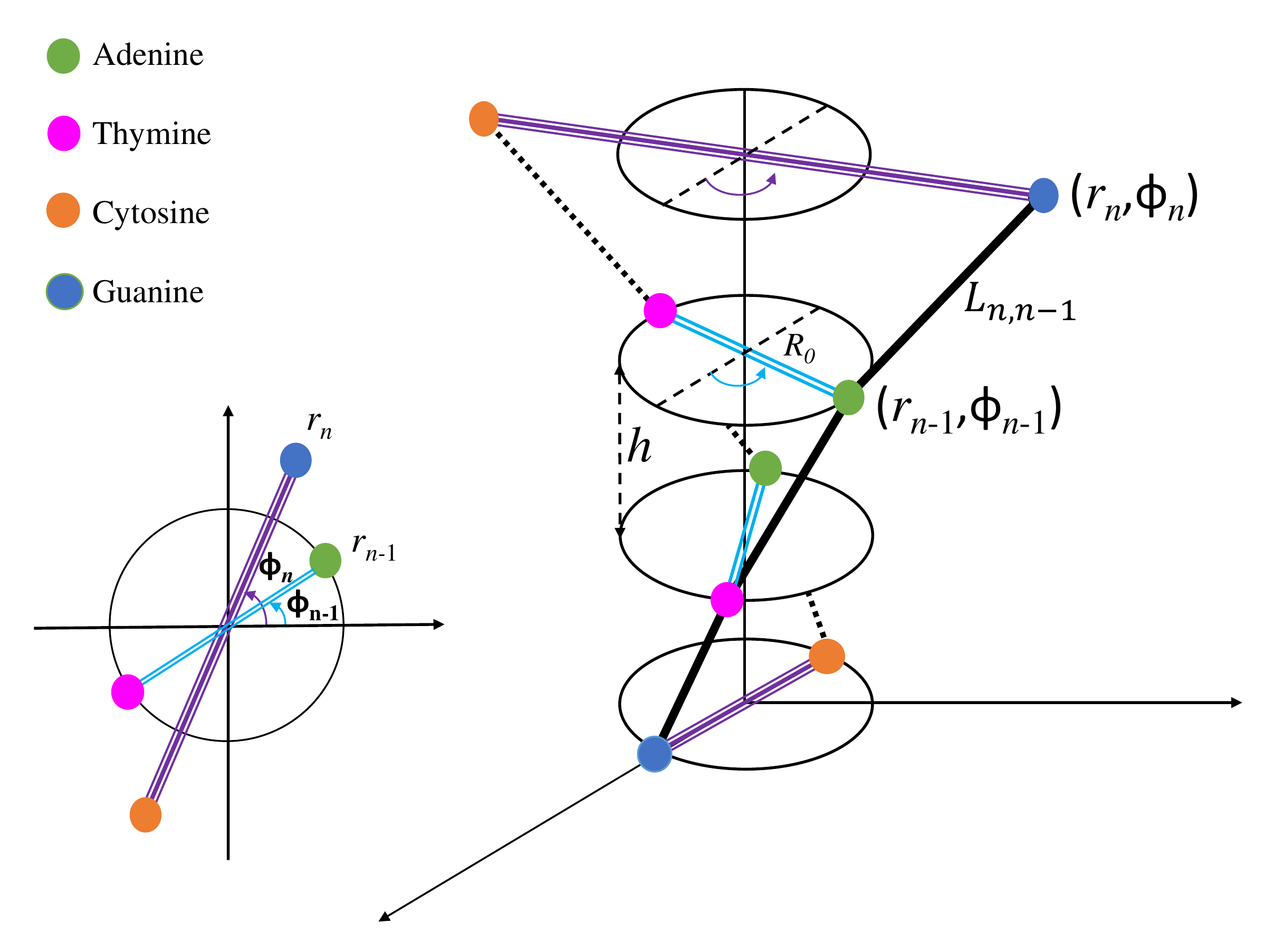}
	\caption{\label{fig:Helix} Schematic representation of the fixed-planes DNA anharmonic twist-opening model. $r_n$ and $\phi_n$ are
radial and torsional degrees of freedom, whose equilibrium values are $R_0$ and $\varphi_\mathrm{eq}$. $L_0$ is the length of the backbone segment connecting the attachment point of the bases along each strand and h is the fixed distance
between two base planes in the B-DNA configuration.}	
\end{figure}

It is worth mentioning that, we consider $\phi_n$ is the same for all base pairs ($\dot{\phi}_n=0,\varphi_n=\varphi$). Now we can derive the equations of motion as
\begin{eqnarray}\label{eq:Langevin}
\frac{d}{dt}\left(\frac{\partial\mathcal{L}}{\partial\dot{r}_n}\right)
= \frac{\partial \mathcal{L}}{\partial r_n}.
\end{eqnarray}
We provide the full analytical expression in the Appendix~\ref{app:EQM}.

The behavior and the evolution in time of a  DNA chain within a cell, is then described by the Langevin equation resulting from Eq.~(\ref{eq:Langevin}) with the addition of a stochastic fluctuating force and a dissipative term on the right-hand-side, in the form
\begin{eqnarray}
-m\gamma \dot {r}_{n}+\sqrt{2\gamma m k_\texttt{B} T}\;\xi_n(t),
\end{eqnarray}
where $\gamma$ is the effective memoryless damping of the system and $\xi(t)$ accounts for thermal noise with the properties $\av{\xi(t)} =0$ and $\av{\xi_m(t)\xi_n(t^{\prime})} =\delta_{mn} \delta(t-t^{\prime})$, where $\av{\cdots}$ denotes the ensemble average. $T$ is the heat bath (cell) temperature and the thermodynamic Boltzmann constant $k_\texttt{B}$. 
In our simulations we used the following values of the parameters, $m=300 amu$, $\gamma=0.5 ps^{-1}$, $D_{AT}=0.05eV$, $\alpha_{AT}={4.2}{\angstrom}^{-1}$,  $K={0.04}{eV/\angstrom^2}$, $\beta={0.5}{\angstrom}^{-1}$ \cite{Barbi2003}. To avoid numerical errors dues to extreme values of the parameters, either very high or very low, we rescale the equations of motion into dimensionless expressions which now takes the form
\begin{eqnarray}\label{eq:ForceLangevin}
\frac{d^2\tilde{r}_n}{d\tau^2} = F_n(\tilde{r}_{n-1},\tilde{r}_n,\tilde{r}_{n+1})-\Gamma \frac{d \tilde{r}_n}{d \tau}+\sqrt{2\Gamma\mathcal{E}} \xi_n(\tau)\,.
\end{eqnarray}
The explicit expression  of $F_n$ is recorded in the Appendix~\ref{app:EQM}. In Eq. (\ref{eq:ForceLangevin}) we introduced the substitutions
\begin{itemize}
	\item The dimensionless displacement $\tilde{r}_n=\alpha r_n$\,,
	\item The dimensionless spring constant $\tilde{K}=\frac{K}{D\alpha^2}$\ and $\tilde{S }=\frac{S}{D\alpha^2}$,
	\item The dimensionless time $\tau=t/t_u$\,,
	\item The dimensionless viscosity coefficient $\Gamma=\gamma t_u$\,,
	\item The dimensionless energy $\mathcal{E}=\frac{k_BT}{D}$\,,
\end{itemize}
where  $\alpha=\alpha_\mathrm{AT}$, $D=D_\mathrm{AT}$ and the characteristic time $t_u=\sqrt{\frac{m}{D\alpha^2}}=0.196 ps$.

\section{ Melting Temperature of twist DNA}\label{sec:Numerical}

We quantify the twisting of DNA by considering the variation of twist angle $\varphi$ around the equilibrium twist angle as $\varphi=\varphi_\mathrm{eq}+\delta$. To numerically calculate the melting transition behavior for different twist angles and heat bath temperatures $T$, we apply the 11-stage symplectic integrator for dissipative systems \cite{Omelyan2003,Harald2000}. We first thermalize the DNA chain very slowly, i.e. the temperature related to the mean kinetic energy, $T_{kin}=(Nk_\texttt{B}/2)^{-1}\sum^{N}_{i=1}\frac{p_i^2}{2m}$  reaches the heat bath temperature. For this we used in the numerical integration fixed boundary conditions. 
After the system has reached the thermal equilibrium state we switch to periodic boundary conditions in order to avoid terminal base pair effects and investigate the denaturation process of DNA as a function of the parameters $T$ and $\delta$. At the equilibrium state we calculate  the mean displacement (average base pair stretching) of $r_i=r_i(t;T,\delta)$ given by
\begin{eqnarray}
\label{eq:MeanY}
\av{r} &=& \frac{1}{n t_s} \sum^{t_s}_{t=1}\sum^{n}_{i=1} r_i
\end{eqnarray}
where $t_s$ is the total simulation time. 
In order to avoid artificial opening of DNA due to finite-size effects, we add to both ends of the chain an extra sequences of 10 $\mathrm{CG}$ bps to harden the boundaries. Accordingly $n$ is smaller than the total number of the DNA base pairs in Eq. (\ref{eq:Lagrangian}). Motivated by the widely studied $\mathrm{P5}$ promoter, we consider after it $n=69$ for our numerical consideration \cite{Kalosakas2004,Alexandrov2009,Falo2013}.

\begin{figure}[htbp]
	\centering
	\subfigure[\label{fig:YMean:a}]{\includegraphics[width=0.49\textwidth]{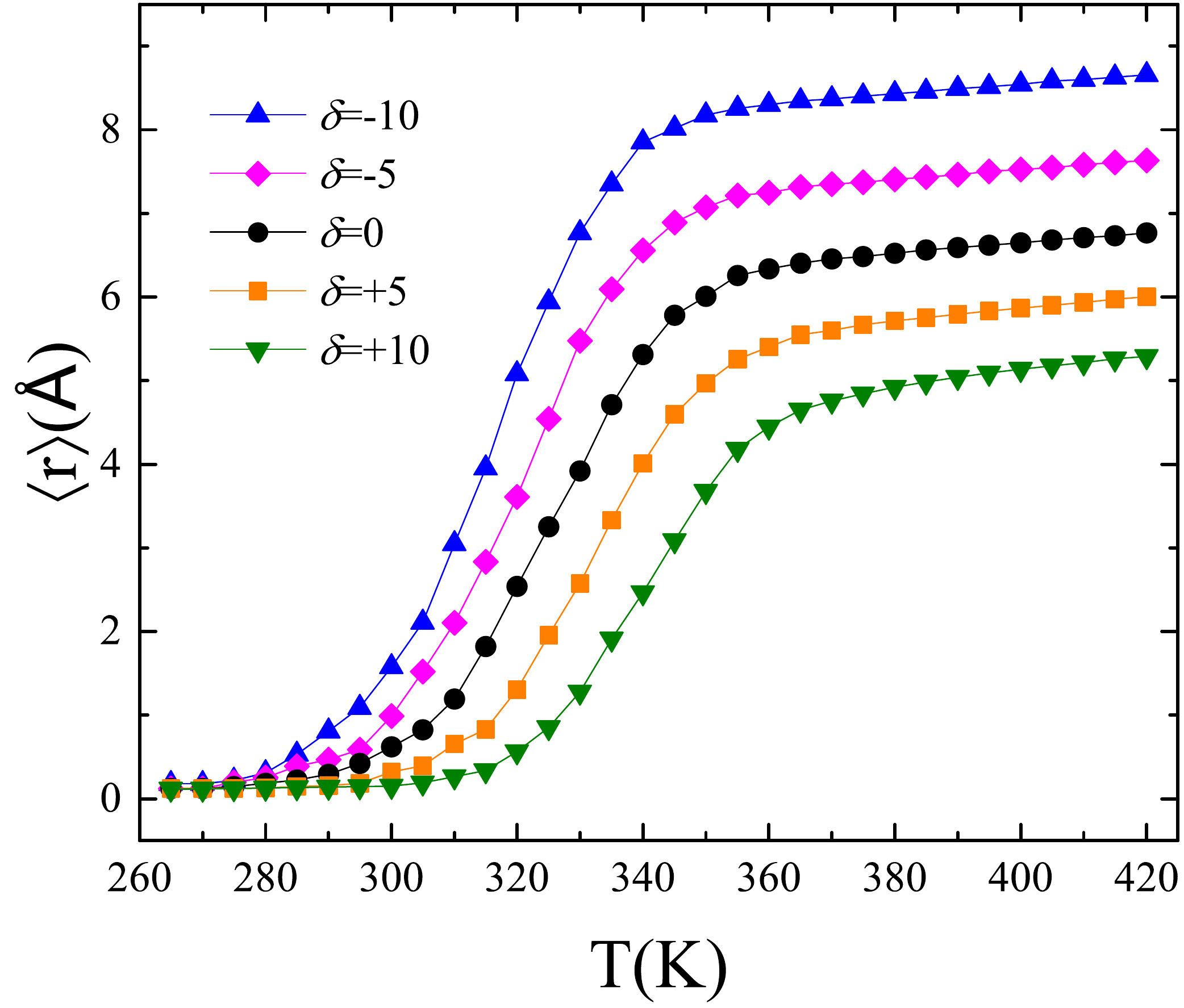}}
	\subfigure[\label{fig:YMean:b}]{\includegraphics[width=0.49\textwidth]{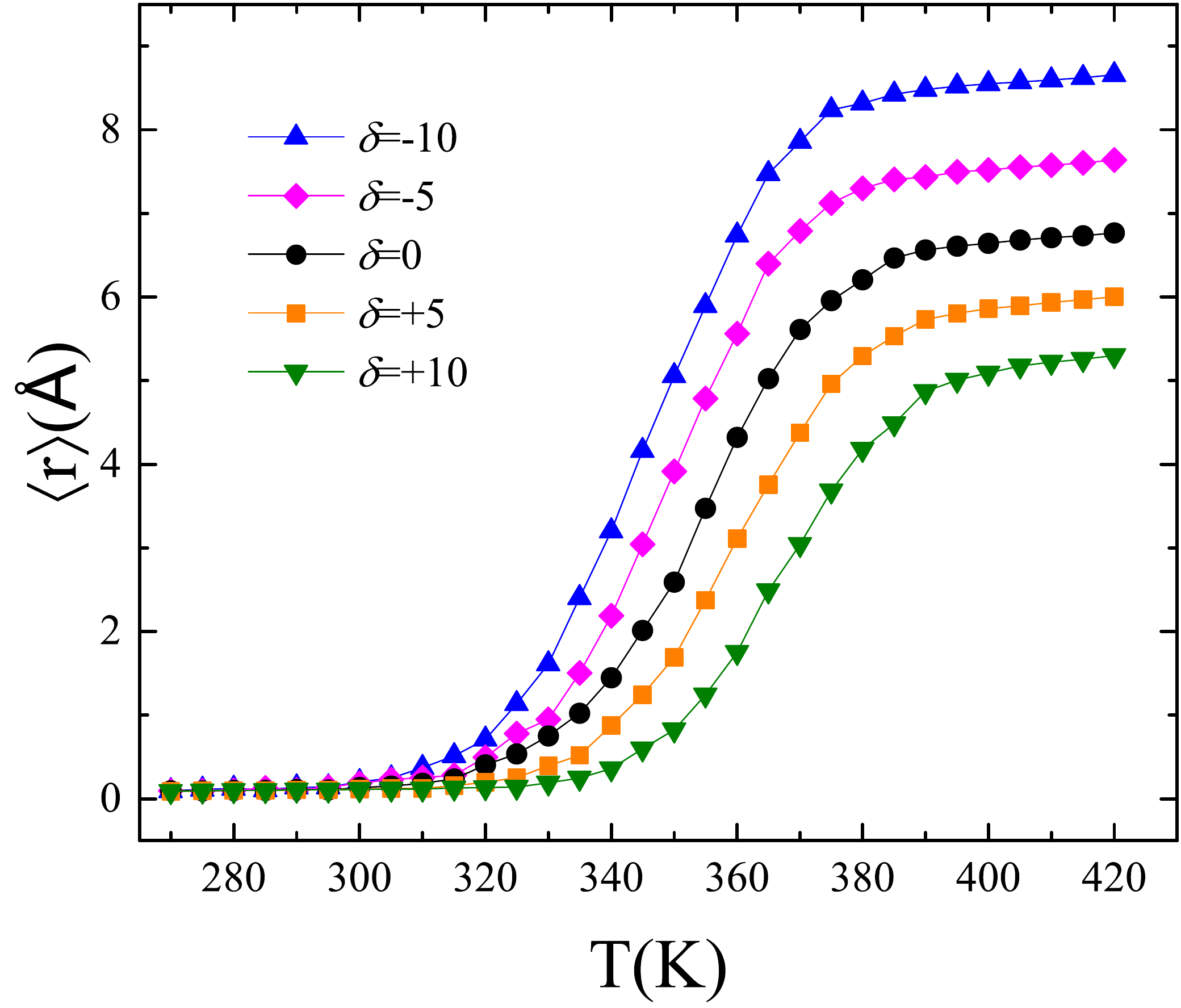}}
	\caption{ \label{fig:YMean} Plot of the mean displacement as function of the equilibrium cell temperature $T$ for different values of the twist angle $\varphi=\varphi_\mathrm{eq}+\delta$  (a) for a homogeneous $\mathrm{AT}$ chain model, and (b) the $\mathrm{P5}$ promoter sequence.}
\end{figure}

The behavior of $\av{r}$ is presented in Figs.~\ref{fig:YMean:a} and \ref{fig:YMean:b} for pure $\mathrm{AT}$ chain and $\mathrm{P5}$ promoter, respectively. The percentage of $\mathrm{AT}$ bps in $\mathrm{P5}$ is $\sim 50.7\% $. We observe in both figures, that higher values of the twist angle correspond to higher resistance of the DNA to the melting process. For the latter to occur one has to considerably  increase the temperature of heat bath. 

\begin{figure}[ht]
	\centering
	\subfigure[\label{fig:MT:b}]{\includegraphics[width=0.49\textwidth]{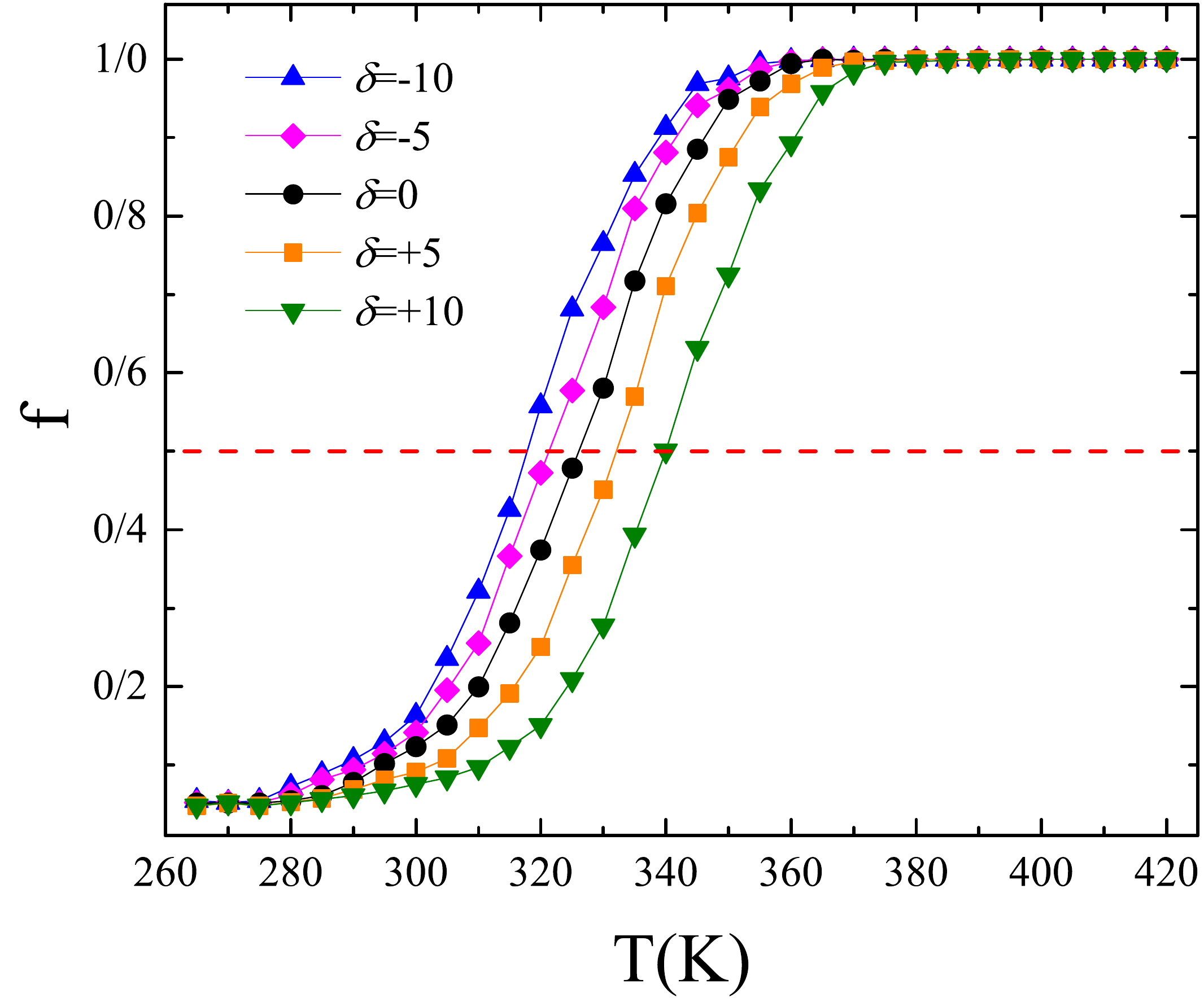}}
	\subfigure[\label{fig:MT:a}]{\includegraphics[width=0.49\textwidth]{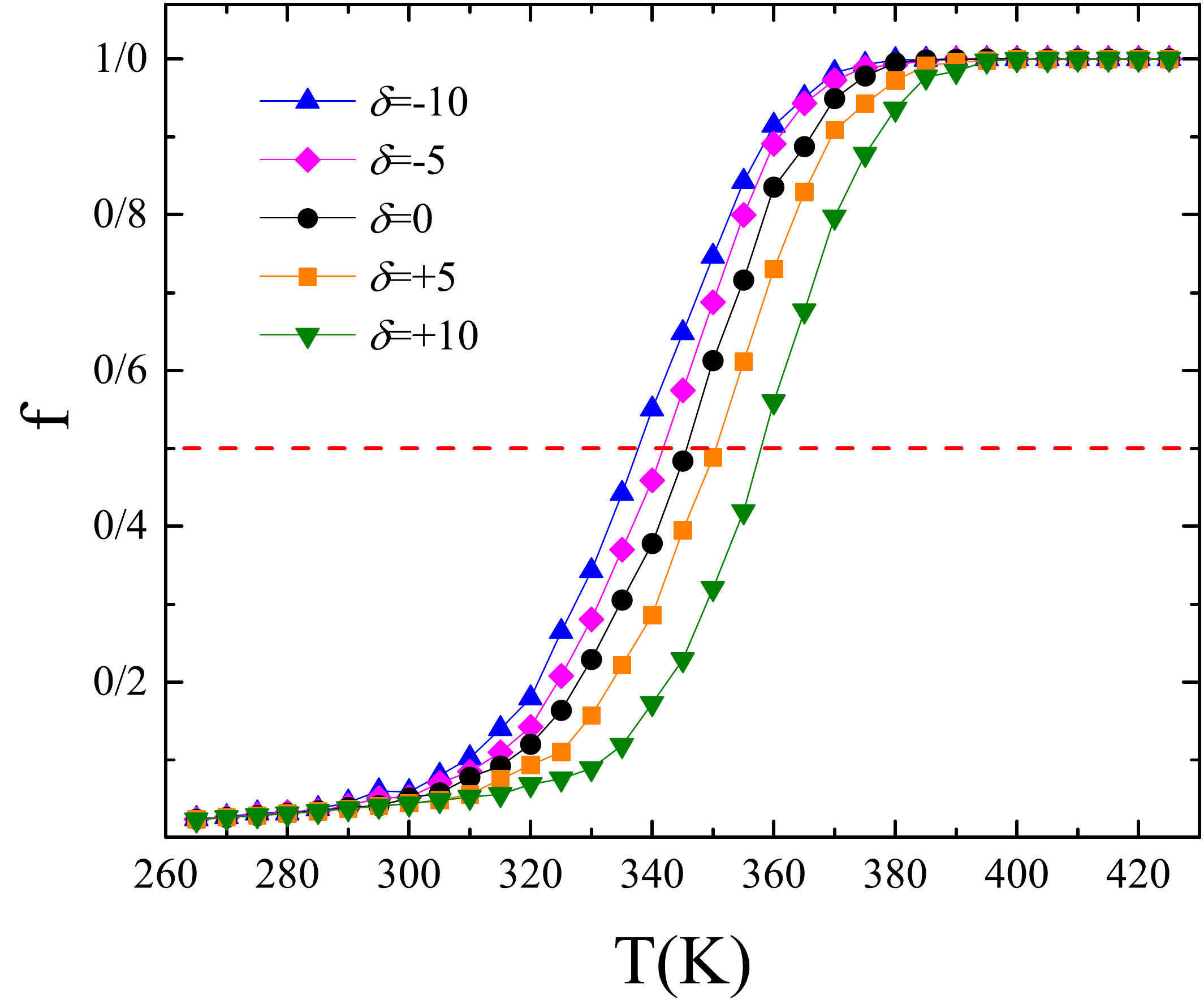}}
	\caption{ \label{fig:MT} The melting transition for homogeneous $\mathrm{AT}$ chain by exploring the fraction of opened base pairs respect to the temperature for different twist angle $\varphi=\varphi_\mathrm{eq}+\delta$.  (a) AT-chain (b) $\mathrm{P5}$ promoter.}
\end{figure}

To quantify the melting process we need to numerically estimate the melting temperature $T_m$. To do so, we first calculate the percentage $f$ of the broken base pairs, which are defined by a vanishing Morse force $F_\mathrm{MP}(r_n)=-\frac{\partial V(r_n)}{\partial r_n}$, i.e. $|F_\mathrm{MP}(r_n-R_0>\Delta r_{tr})|\to0$, for various reservoir temperatures.
The former condition is satisfied by the threshold displacement value $\Delta r_{tr}={1}{\angstrom}$ signaling the transverse opening of DNA.
Then, we plot in Fig.~\ref{fig:MT:b} $f$ as a function of temperature for different twist angles. The temperature in which the half of base pairs are broken, is identified as the melting temperature $T_m$ (red dash line in Fig.~\ref{fig:MT:b}) \cite{Falo2010}. 

\begin{figure}[hbtp]
	\centering
	\includegraphics[width=0.5\textwidth]{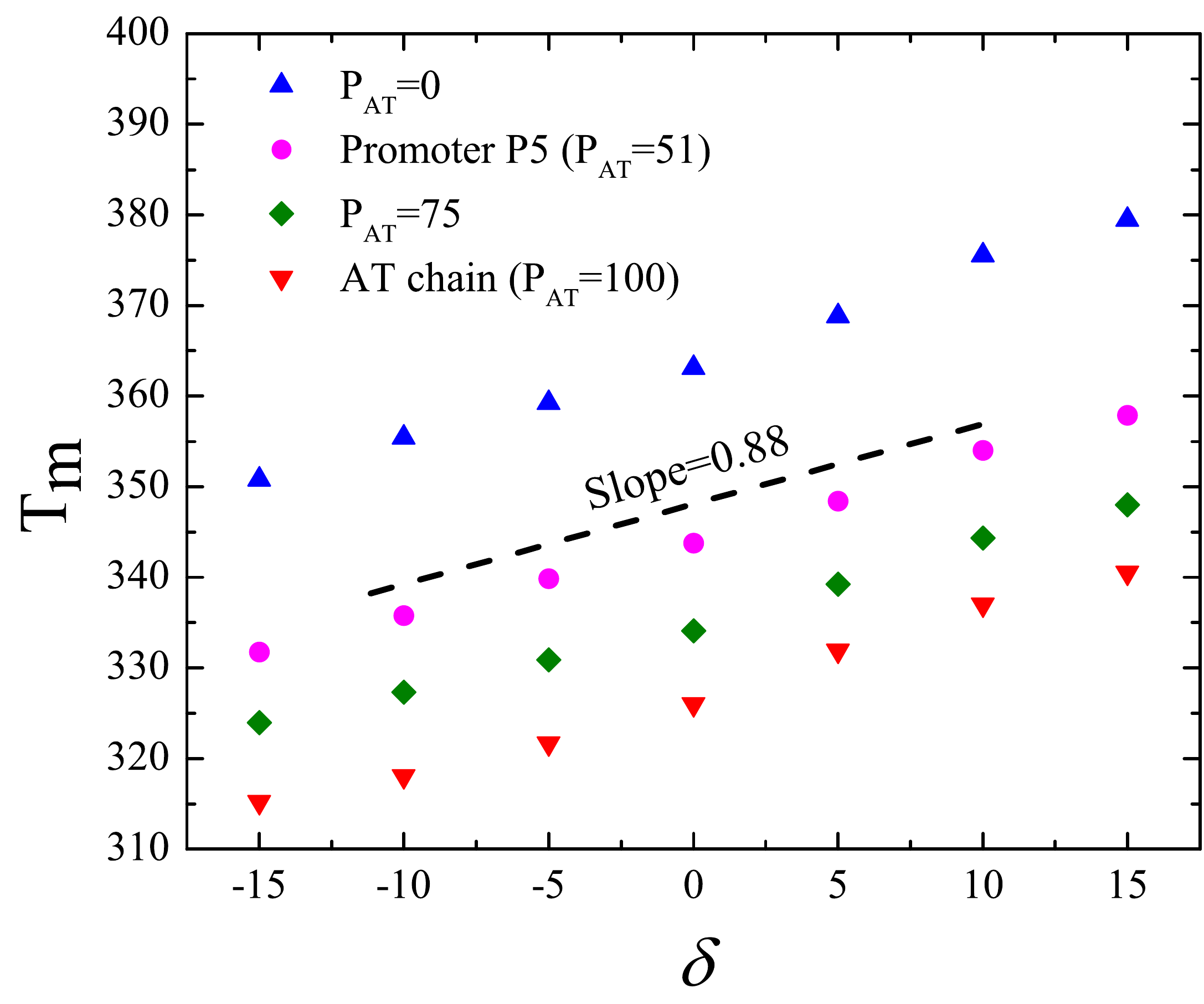}
	\caption{\label{fig:melting} Melting temperature as a function of the twist angle for various percentages of $\mathrm{AT}$ bps. The $\mathrm{P5}$ promoter contains $50.7\%$ of $\mathrm{AT}$ bps.}	
\end{figure}

In Fig.~\ref{fig:melting} we plot the relation between the calculated melting temperature $T_m$ and the twist angle $\delta$ for different chains. Where $P_{AT}$ is the percentage of $\mathrm{AT}$ base pairs in the chain, i. e. $P_{AT}=100$ means a pure $\mathrm{AT}$ chain. As we expected, by increasing of $\delta$ melting temperature increase for all kind of the chains. We can see linear behavior of melting temperature as a function of $\delta$. Also, in equilibrium twist angle ($\delta=0^\circ$), the melting temperature coincide with equation $T_m=365-0.4P_{AT}$ \cite{Haris2019,Kalosakas2009,Skokos2020}. For example, in agreement with the former relation, melting temperature of homogeneous $\mathrm{AT}$ chain in unfolded case has been estimated to be around $T_m=326.2\mathrm{K}$ \cite{Boian2009,Wells1970}. 
For a quantitative description, we fit the curves in Fig. \ref{fig:melting} with a linear function, i.e., $T_m(\delta)=A+B \delta$, and record for each case the coefficients $A$ and $B$ in Table \ref{table1}. The goodness of the linear fit is given by the adjusted $R^2\in[0,1]$ statistical index. We observe that for all  $\delta$ the variation of the slope of $T_m(\delta)$ due to the different percentage of $\mathrm{AT}$ in the sequence is confined in a narrow interval corresponding to $0.82< B< 0.97$.
\begin{table}[h]
	\begin{tabular}{c||cccc}
	$P_{AT}(\%)\to$	& 0 & 51 & 75 & 100 \\ \hline \hline
	$A$	& 364.6  & 344.5 & 335.4 & 326.2 \\ \hline
	$B$	& 0.97 & 0.88 & 0.82 & 0.89 \\ \hline
	adj. $R^2$	& 0.998 & 0.997 & 0.998 & 0.998 \\ 
	\end{tabular}
\caption{Parameter values of the fitting function $T_m(\delta)=A+B \delta$ for the curves in Fig. \ref{fig:melting}.}
\label{table1}
\end{table}

\begin{figure*}[htbp]
	\centering
	\subfigure[\label{fig:Bubble:a}]{\includegraphics[width=0.75\textwidth]{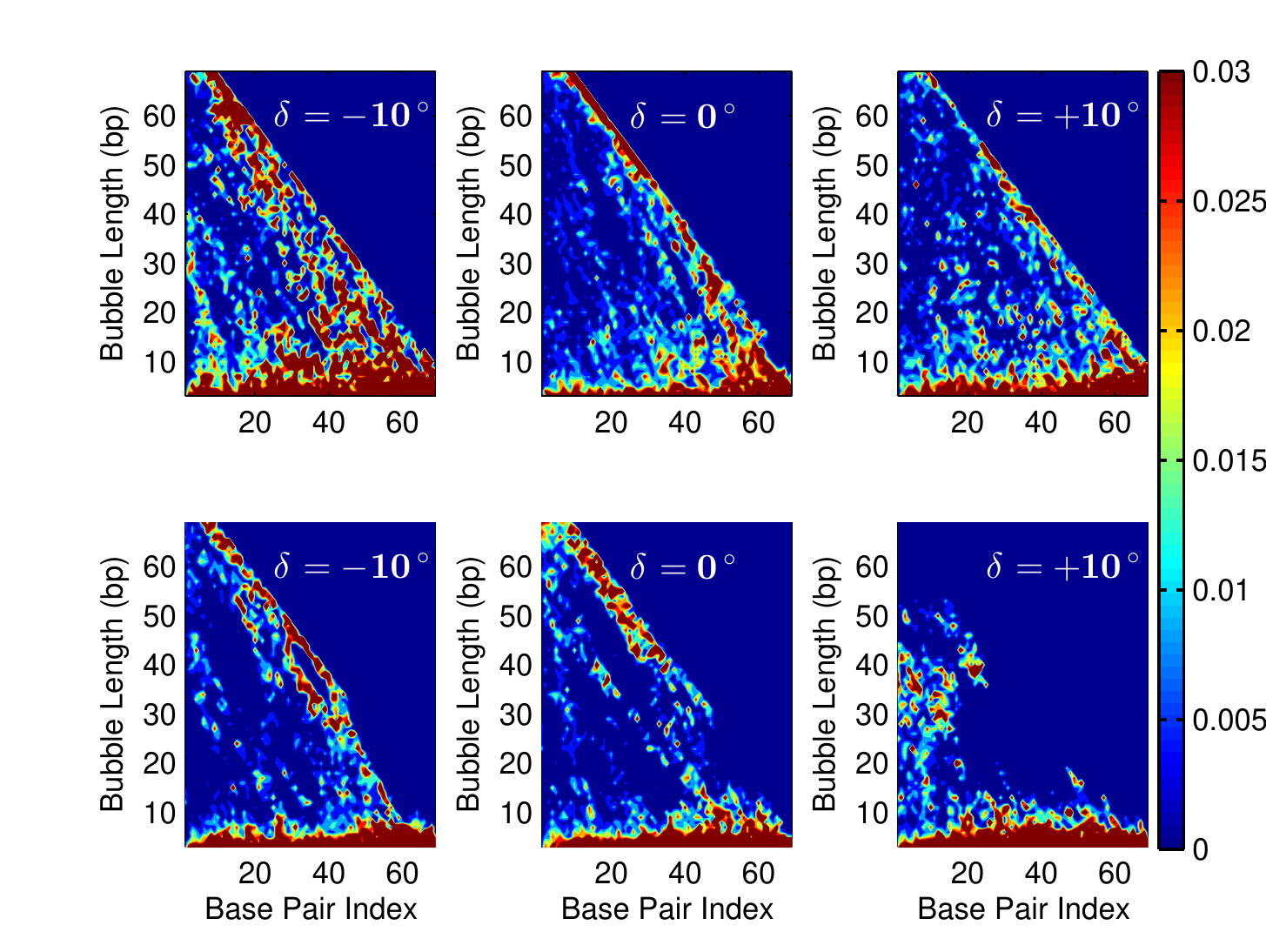}}
	\subfigure[\label{fig:Bubble:b}]{\includegraphics[width=0.75\textwidth]{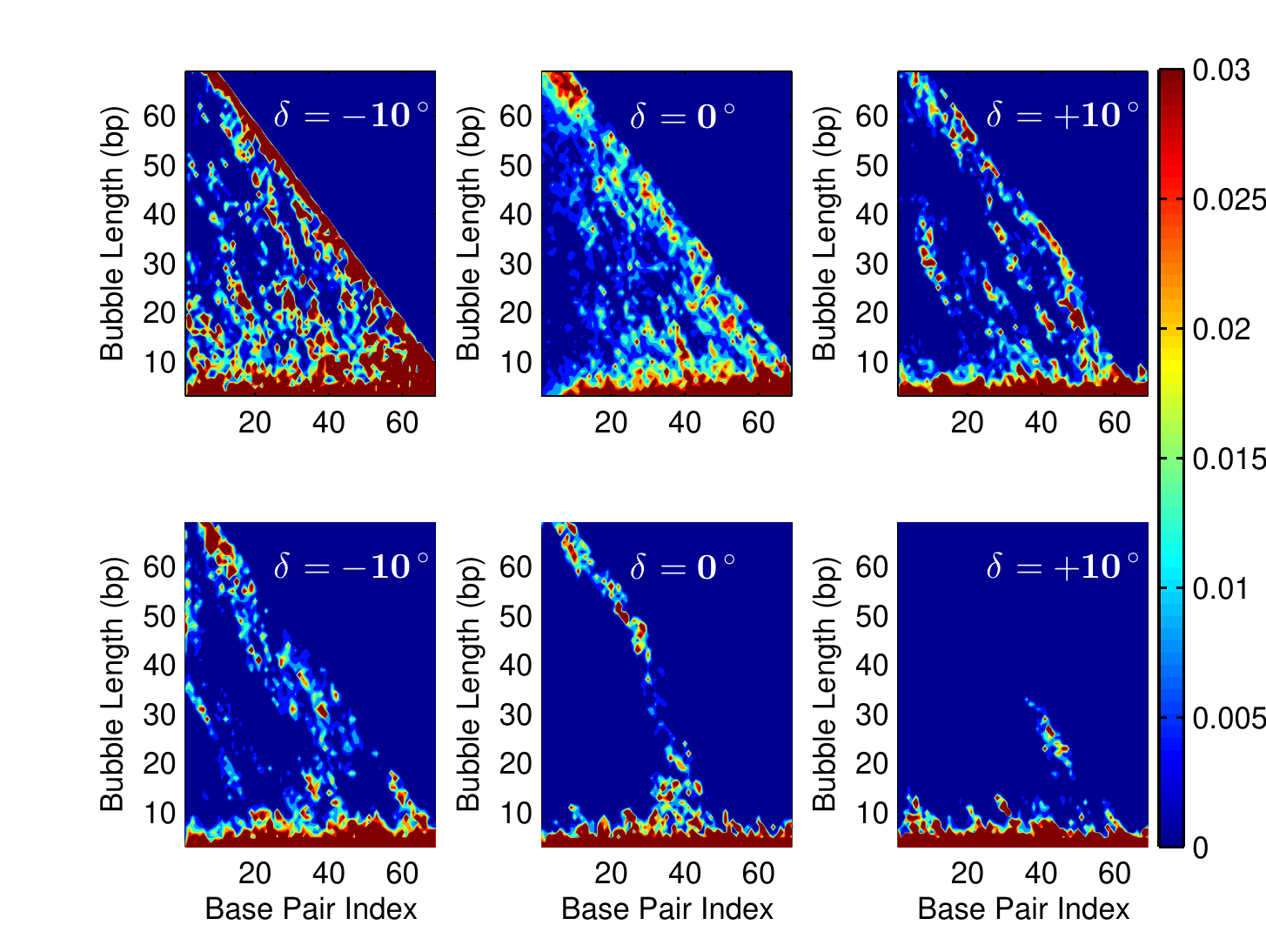}}
	\caption{ \label{fig:Bubble} Probability of opening distribution $P_n(\ell)$ in different twist angle $\delta=-10^\circ,\,0^\circ,\, +10^\circ$ for (a) homogeneous $\mathrm{AT}$ chain and (b) for $\mathrm{P5}$ promoter chain. The probability of bubble formation are given by the color scale. In both case, the top rows illustrate the simulation of bubble formation at $T>T_m$ while the bottom rows are at $T<T_m$.}  
\end{figure*}

\section{Bubbles formation}\label{sec:BF}

The creation of a bubble and its lifetime is very important in the dynamics of double-strand DNA. This stochastic process which could in principle affect processes like transcription or replication is most appropriately described in terms of a probability. Therefore, we calculate the probability of bubble existence of length $\ell$ that begins its formation at the $n$th base pair as
\begin{eqnarray} \label{eq:Probablity}
P_n(\ell)=\frac{\sum_{k=1}^{M}\left[\sum_{j=1}^{t_s}\Delta t_j^k(\ell)\right]}{\sum_{\ell=3}^{69-n}\sum_{k=1}^{M}\left[\sum_{j=1}^{t_s}\Delta t_j^k(\ell)\right]}\,,
\end{eqnarray}
where $\Delta t_j^k(\ell)$ is the life time of the double-strand separation of amplitude bigger than $y_0$, spanning $\ell>3$  in consecutive base pairs beginning at the $n$th base pair in the $k$th simulation. Our averaging is over $M= 1000$ simulations. We use again fixed boundary conditions to avoid the complete opening of the chain. There are many aspects that can be studied based on $P_n(\ell)$, e.g.,  the dependence of bubble formation on the internal nucleotide structure, i.e., the patterns of the $\mathrm{AT}/\mathrm{CG}$ repetitions \cite{Alexandrov2006,Alexandrov2009}.
Here,  we are particularly interested on the effect of DNA twisting on the bubble generation. 

In Figs.~\ref{fig:Bubble} the colormaps of the probability $P_n(\ell)$ in Eq.~(\ref{eq:Probablity}) are presented for the $\mathrm{AT}$ chain and $\mathrm{\mathrm{P5}}$ promoter, respectively,  as a function of $\ell$ and $n$ for four representative twist angles $\delta$. Particularly, in Figs. \ref{fig:Bubble:a}, \ref{fig:Bubble:b} the top rows are depict the simulation of bubble formation  at $T>T_m$ while the bottom rows are at $T<T_m$. The 10 bps at the beginning and the end of the sequence are not included in the figures. The most dominant and consistent in all cases effect of the twist angle $ \delta$ is that the probability of having big bubble formation substantially decreases as we increase $\delta$. On the other hand, the formation of small bubbles is almost sure for all angles with  $\delta\in[-10^\circ, +10^\circ]$.
The diagonal separation between zero and non-zero probabilities for the unfolded sequences is related to their finite size.
Regarding the temperature dependence we observe for $T>T_{m}$ a relative shift of small bubbles to big bubbles along the diagonal,  compared to $T<T_{m}$. This is expected and understandable from the thermodynamic point of view since the sites become more excited.
Last but not least, a comparison between  $\mathrm{P5}$ promoter and the pure $\mathrm{AT}$ chain, for all twist angles, unveils a pattern of smaller formation probability values in the former case. This can is explained by the fact that $\mathrm{CG}$ bonds existing in the $\mathrm{P5}$ promoter are stronger than the $\mathrm{AT}$ bonds.\\

\begin{figure}[hbtp]
\centering
\includegraphics[width=0.6\textwidth]{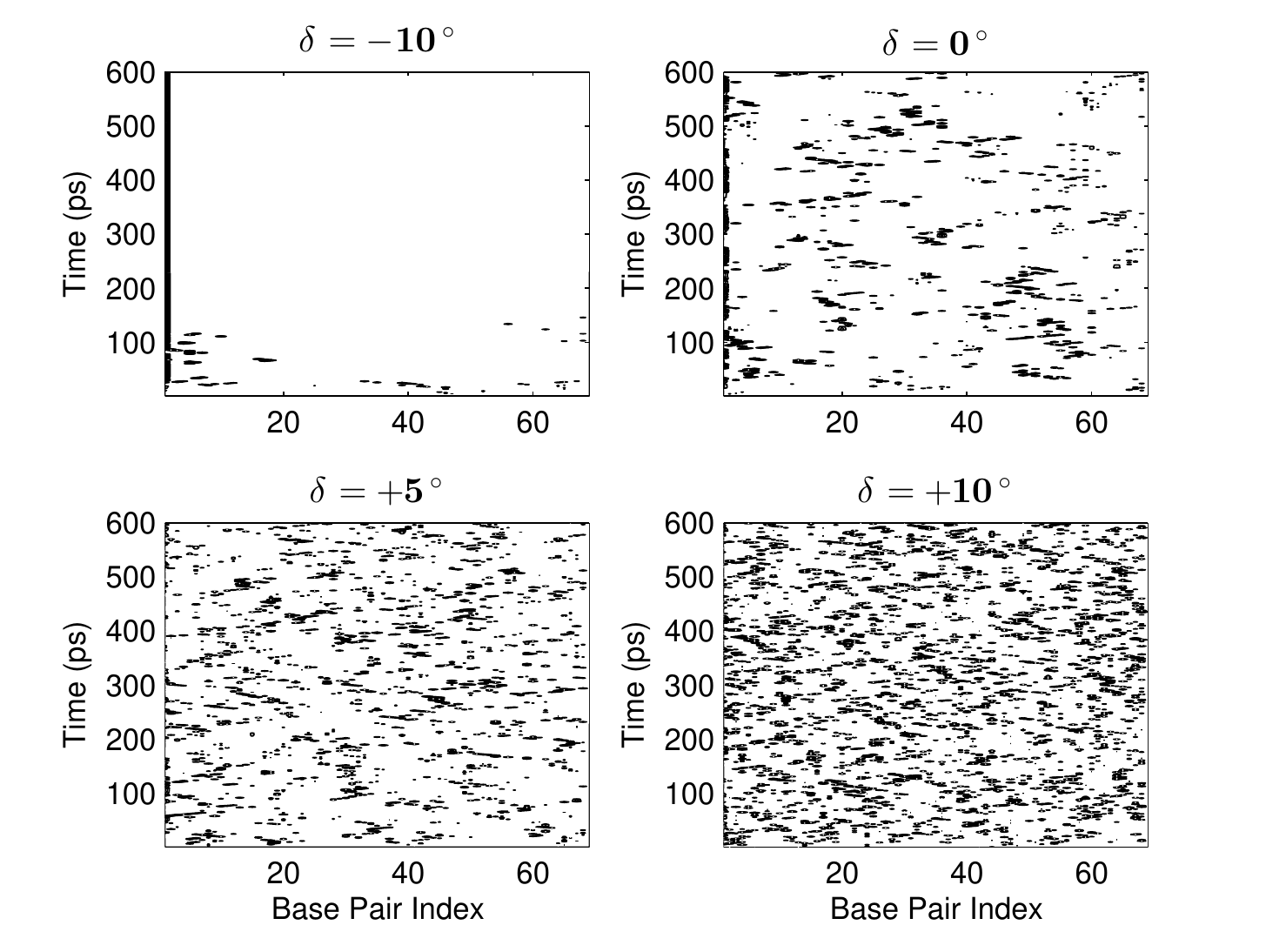}
\caption{\label{fig:trajectory} Typical molecular dynamics simulation trajectories for the $\mathrm{P5}$ promoter	sequence at temperature T = 350 K and twist angles $\delta=-10^\circ,\,0^\circ,\, 5^\circ,\,+10^\circ$. The horizontal axis extends along $\mathrm{P5}$ promoter bps and the vertical axis gives to time in $ps$ scale. Trajectory time is $600 ps$. Dark areas correspond to open base pairs. Long-living bubbles are clearly observed when $\delta=-10^\circ$. For better view, we plotted the figures a little before zero.} 
\end{figure}

Another interesting aspect of the bubble formation phenomenon is its stability in time, which plays a pivotal role in the understanding in biofunctional operations of DNA molecules. 
Our model unveils that the degree of twisting in a DNA sequence significantly affects the lifetime $\Delta t$ of bubbles exhibiting a monotone decreasing behavior between $\delta$ and $\Delta t$. 
In Fig.~\ref{fig:trajectory} the amplitude of the base pair stretching is recorded in a binary code for $T=350 K$, which is slightly above the denaturation temperature of $\mathrm{P5}$ promoter. The white and black colors correspond to fully closed and open base pairs, respectively. For $\delta=-10^\circ$, after approximaetly 100$ps$ there is one big bubble in the size of the chain. This means that promoter is completly opened. Such regions correspond to the ``denaturation event'' observed in the experiments. By increasing the twist angle, the continuously DNA opening is interupted while more small black spots are detected for limited time. These black spots correspond to regions with small base pair stretching, over a few consecutive bases. In other words, higher values of $\delta$ creates instabilities in the DNA opening localized in position and time. Here, again, and in agreement with our preceding result, we observe that for higher twist angles DNA becomes more rigid against opening with small bubble amplitudes.

\section{ Summary and Discussion}\label{sec:Dis}

We studied the effect of a heat bath on the dynamics of a DNA molecule using  the modified $\mathrm{BCP}$ model by considering twist angle $\varphi$ between two consecutive base pairs in the stacking interaction potential. Also, we added dissipative and thermal fluctuations terms to study more realistic condition of DNA dynamics.   
Specifically, we analyzed the melting transition of a thermalized DNA sequence under scrutiny.  For this, we calculated  the melting temperature and studied its behavior as a function of $\varphi\in[\varphi_\mathrm{eq}-10^\circ,\varphi_\mathrm{eq}+10^\circ]$ for various percentages of the $\mathrm{AT}$ content in the sequence varying from $0\%$ to $100\%$ (including the $\mathrm{P5}$ promotor with $51\%$ of $\mathrm{AT}$ repetitions). In all cases, the melting temperature $T_m$ was behaving linearly to variation of twist angle. The slope related factor of the line is confined in an narrow range of values, $[0.82,0.97]$, with the tendency to increase for higher $\mathrm{AT}$ percentages. 
The equilibrium angle for which DNA is most stable behavior against thermal disruption of the base pair bonds, is determined in literature to be $\varphi\approx34.6^\circ$. In accordance to this, our analysis revealed that for $\varphi\geq\varphi_\mathrm{eq}$ the melting temperature exceeds $345K$ which piratically means that DNA reaches locally to high temperature before denaturation. Moreover, as shown in Fig.~\ref{fig:melting}, DNA needs higher temperature for melting. Accordingly, we can say that twisted DNAs more than $\varphi_\mathrm{eq}$ do not undergo the melting transition. 

Next, considering a homogeneous $\mathrm{AT}$ chain and the $\mathrm{P5}$ promoter, we studied for four representative twist angles, $\{\varphi_\mathrm{eq}-10^\circ,\varphi_\mathrm{eq},\varphi_\mathrm{eq}+10^\circ\}$, the probability $P_n(\ell)$ of a bubble formation as a function of its length $\ell$ and the position $n$ of the starting base pair. We presented the results in a colormap.
The common feature for both sequences is that the highest probability values for observing a big bubble formation is recorded in the case the less folded DNA ($\delta=-10^\circ$). Towards the equilibrium angle bubbles of small length were more probable. For $\delta=+10^\circ$ the bubble length was almost uniformly distributed (small fluctuations around an average $\ell$) among all base pairs with $\ell(\varphi_\mathrm{eq}) < \ell(\varphi<\varphi_\mathrm{eq})$. The difference of the two sequences in the percentage of the $\mathrm{AT}$ content did not qualitatively change the probabilistic picture in the $(\ell,n)$ plane. The stability of the bubble formation in time as a function of the twist angle, is recorded in Fig. \ref{fig:trajectory} for a representative temperature $T=350K$. We observed that in $\varphi<\varphi_\mathrm{eq}$ case the DNA chain is completely opened and maintains this state for long times. Again, the behavior changes consistently by increasing $\varphi$, namely bubbles are created locally (a few bps) for very short times.

Finally, it would be of interest to  investigate the current model by including  a solvation barrier in the on-site potential \cite{Falo2010}, to find out whether and how the barrier affects the melting temperature and the bubble formation with respect to the twist angle.

\section*{Acknowledgements}
The authors acknowledge the ORAU grant with PN 17098 and the state-targeted program ``Center of Excellence for Fundamental and Applied Physics" (BR05236454) by the Ministry of Education and Science of the Republic of Kazakhstan.

\appendix*
\section{ Rescaled Equation of motion}\label{app:EQM}

The full analytical expression in Eq. (\ref{eq:Langevin}) is given by
\begin{eqnarray}\label{App1}
\nonumber
m \ddot r_{n} &=& 2\alpha_n D_n\left(e^{-\alpha_n (r_n-R_0)}-1\right)e^{-\alpha_n (r_n-R_0)}\nonumber \\
&&-2K\left[(L_{n,n-1}-L_0)\frac{r_n-r_{n-1}\cos\varphi}{L_{n,n-1}}+(L_{n+1,n}-L_0)\frac{r_n-r_{n+1}\cos\varphi}{L_{n+1,n}}\right] \nonumber \\
&&+Se^{-\beta(r_n+r_{n-1}-2R_0)}(r_n-r_{n-1})[\beta(r_n-r_{n-1})-2]\nonumber \\
&&+Se^{-\beta(r_{n+1}+r_n-2R_0)}(r_{n+1}-r_n)[\beta(r_{n+1}-r_n)+2].
\end{eqnarray}
Introducing the dimensionless stretching of the base pairs as $\tilde{r}_n=\alpha r_n$, $\tilde{R}_0=\alpha R_0$ and substituting $b=\frac{\beta}{\alpha}$ and $a_n=\frac{\alpha_n}{\alpha}$, we can rewrite the equation of motion as
\begin{eqnarray}\label{eq:Langevin2}
m \frac{d^2\tilde{r}_n}{dt^2} &=& 2\alpha\alpha_n D_n\left(e^{-a_n (\tilde{r}_n-\tilde{R}_0)}-1\right)e^{-a_n (\tilde{r}_n-\tilde{R}_0)}\nonumber \\
&&-2K\left[(\tilde{L}_{n,n-1}-\tilde{L}_0)\frac{\tilde{r}_n-\tilde{r}_{n-1}\cos\varphi}{\tilde{L}_{n,n-1}}+(\tilde{L}_{n+1,n}-\tilde{L}_0)\frac{\tilde{r}_n-\tilde{r}_{n+1}\cos\varphi}{\tilde{L}_{n+1,n}}\right] \nonumber \\
&&+Se^{-b(\tilde{r}_n+\tilde{r}_{n-1}-2\tilde{R}_0)}(\tilde{r}_n-\tilde{r}_{n-1})[b(\tilde{r}_n-\tilde{r}_{n-1})-2]\nonumber \\
&&+Se^{-b(\tilde{r}_{n+1}+\tilde{r}_n-2\tilde{R}_0)}(\tilde{r}_{n+1}-\tilde{r}_n)[b(\tilde{r}_{n+1}-\tilde{r}_n)+2]\nonumber \\
&&-m\gamma \frac{d\tilde{r}_n}{dt}+\alpha\sqrt{2\gamma m k_\texttt{B} T}\;\xi_n(t).
\end{eqnarray}
Next, we introduce the dimensionless time $\tau=\sqrt{\frac{D\alpha^2}{m}}t$ and the substitutions $\lambda_n=\frac{D_n\alpha_n}{D\alpha}$, so that
\begin{eqnarray}\label{eq:Langevin3}
 \frac{d^2\tilde{r}_n}{d\tau^2} &=& 2\lambda_n\left(e^{-a_n (\tilde{r}_n-\tilde{R}_0)}-1\right)e^{-a_n (\tilde{r}_n-\tilde{R}_0)}\nonumber \\
&&-2\frac{K}{D\alpha^2}\left[(\tilde{L}_{n,n-1}-\tilde{L}_0)\frac{\tilde{r}_n-\tilde{r}_{n-1}\cos\varphi}{\tilde{L}_{n,n-1}}+(\tilde{L}_{n+1,n}-\tilde{L}_0)\frac{\tilde{r}_n-\tilde{r}_{n+1}\cos\varphi}{\tilde{L}_{n+1,n}}\right] \nonumber \\
&&+\frac{S}{D\alpha^2}e^{-b(\tilde{r}_n+\tilde{r}_{n-1}-2\tilde{R}_0)}(\tilde{r}_n-\tilde{r}_{n-1})[b(\tilde{r}_n-\tilde{r}_{n-1})-2]\nonumber \\
&&+\frac{S}{D\alpha^2}e^{-b(\tilde{r}_{n+1}+\tilde{r}_n-2\tilde{R}_0)}(\tilde{r}_{n+1}-\tilde{r}_n)[b(\tilde{r}_{n+1}-\tilde{r}_n)+2]\nonumber \\
&&-\gamma\sqrt{\frac{m}{D\alpha^2}}\frac{dr_n}{d\tau}+\frac{\sqrt{2\gamma m k_BT}}{D\alpha}\xi_n\left(\sqrt{\frac{m}{D\alpha^2}}\tau\right).
\end{eqnarray}
Finally, in the former equation we rewrite the noise term as
\begin{eqnarray}\label{App04}
\xi\left(\sqrt{\frac{m}{D\alpha^2}}\tau\right)\rightarrow\sqrt[4]{\frac{D\alpha^2}{m}}\xi(\tau)\,,
\end{eqnarray}
which is justified due to Dirac delta function and Gaussian noise properties, $\av{ \xi_n(A\tau)\xi_n(A\tau^{\prime}) } = A^{-1} \delta(\tau-\tau^{\prime})$ and $\xi(A\tau)\rightarrow\frac{1}{\sqrt{A}}\xi(\tau)$, respectively.

By considering
\begin{eqnarray}\label{eq:Force}
F_n(\tilde{r}_{n-1},\tilde{r}_n,\tilde{r}_{n+1}) &=& 2\lambda_n\left(e^{-a_n (\tilde{r}_n-\tilde{R}_0)}-1\right)e^{-a_n (\tilde{r}_n-\tilde{R}_0)}\nonumber \\
&&-2\tilde{K}\left[(\tilde{L}_{n,n-1}-\tilde{L}_0)\frac{\tilde{r}_n-\tilde{r}_{n-1} \cos\varphi}{\tilde{L}_{n,n-1}}+(\tilde{L}_{n+1,n}-\tilde{L}_0)\frac{\tilde{r}_n-\tilde{r}_{n+1} \cos\varphi}{\tilde{L}_{n+1,n}}\right] \nonumber \\
&&+\tilde{S}e^{-b(\tilde{r}_n+\tilde{r}_{n-1}-2\tilde{R}_0)}(\tilde{r}_n-\tilde{r}_{n-1})[b(\tilde{r}_n-\tilde{r}_{n-1})-2]\nonumber \\
&&+\tilde{S}e^{-b(\tilde{r}_{n+1}+\tilde{r}_n-2\tilde{R}_0)}(\tilde{r}_{n+1}-\tilde{r}_n)[b(\tilde{r}_{n+1}-\tilde{r}_n)+2]\,,
\end{eqnarray} 
and substituting Eq. (\ref{App04}) into Eq. (\ref{eq:Langevin3}) we are led to  Eq. (\ref{eq:ForceLangevin}) in the text, namely
\begin{eqnarray*}\label{eq:RescaleLangevin}
\frac{d^2\tilde{r}_n}{d\tau^2} = F_n(\tilde{r}_{n-1},\tilde{r}_n,\tilde{r}_{n+1})-\Gamma \frac{d \tilde{r}_n}{d \tau}+\sqrt{2\Gamma\mathcal{E}} \xi_n(\tau)\,.
\end{eqnarray*}
%


\end{document}